\documentclass[page-classic]{epl2}
\usepackage{graphicx}
\usepackage{amsmath}
\usepackage{amssymb}

\title{Topological glass states}

\author{Tai-Kai Ng\inst{1}\and Yi Zhou\inst{2} \and Lei-Han Tang\inst{3}}
\shortauthor{T.-K. Ng \etal}

\institute{
\inst{1} Department of Physics, Hong Kong University of Science and
Technology, Clear Water Bay Road, Kowloon, Hong Kong\\
\inst{2} Department of Physics, University of Hong Kong, Pok Fu Lam Road,
Hong Kong\\
\inst{3} Department of Physics, Hong Kong Baptist University, Kowloon Tong, Kowloon,
Hong Kong}
\date{ \today }


\abstract{
In connection with recent discussion of topological order and topological phase
transitions in quantum systems, we reexamine circumstances that lead to the
appearance of a topological glass in certain classical lattice spin models.
Local bonding enforces constraints on low energy states which organize themselves
into topologically distinct classes that break ergodicity but not any apparent
symmetry as in the usual Landau theory of phase transitions.
Various properties of such a topological glass are demonstrated
using two classical Ising-like models.
}

\begin{document}

\maketitle

Quantum field theories with non-trivial topological structures have attracted a
great deal of interest in recent years\cite{fqhe,wen,fulde,kitaev,tao}.
Although the fundamental variables
of these systems are topologically simple (e.g., ordinary spins or boson/fermion
fields on a periodic lattice), their low energy properties are characterized
by a non-trivial topological order and order parameters that are non-local
functions of the original microscopic variables. The effective variables may take
the form of fermions or bosons dressed by Chern-Simon field\cite{fqhe} or
string-like objects\cite{wen, fulde}.
Topological excitations with fractionalized charge (or spin) quantum number often
appear in these systems as a result of the topological order. The study of this
class of problems is important for a better understanding of some of the more
exotic states of matter such as the FQH states.

Many existing examples of topological order are from quantum systems with a
non-trivial Berry phase structure. Non-trivial topology may also arise, however,
from the spatial bonding pattern in the ground state which can be understood
from a purely classical viewpoint. Using simple and clear examples from
statistical mechanics, we examine in this paper how various aspects of
topological order, such as ground state degeneracy, reduced phase space,
topological excitations and topological phase transitions, come together
in such a scenario. Whereas topological order is often characterized
by topological excitations with fractionalized charge (or spin)
quantum number in quantum systems, it manifests itself as breaking of ergodicity
in classical systems. These related but also distinct features
are noteworthy as they contribute to a more complete identification and
classification of topologically ordered many-body quantum states.

The classical systems we consider have an extensive ground state entropy
and no apparent spatial order in the usual sense of the word at any temperature.
Nevertheless, the low-energy part of the phase space decomposes into
topologically distinct sectors, leading to the breaking of ergodicity
at sufficiently low temperatures.
In this respect, they resemble the glass state in disordered systems, though the
energetics and dynamics of excitations that restore ergodicity are expected to
be quite different in the two cases. This point will also be discussed
in some detail below.

\section{The AKLT state}
It is easiest to illustrate the concept of a topological
glass with concrete Hamiltonians. We first consider an example which connects
topological glass states in classical systems and topological
order in quantum systems. The example is a classical analogue of the
Affleck-Kennedy-Lieb-Tasaki (AKLT) Valence Bond Solid state\cite{aklt} in
quantum spin chains and is defined by the following classical one-dimensional Hamiltonian,
\begin{equation}
H_{\rm s}=J\sum_{i={\rm odd\ integer}}\sigma_i\sigma_{i+1},
\label{H_s_1D}
\end{equation}
where $J>0$ and $\sigma_i=\pm 1/2$ are classical Ising spins (see
Fig. 1). Since the Hamiltonian is a sum of disjoint terms,
$\langle \sigma_i\rangle=0$ and there is no long-distance
correlation in the system at any temperature. Now let us introduce
the composite spin variables $S_j=\sigma_{2j}+\sigma_{2j+1}$ which
by construction have possible values $S_j=-1,0,1$. There is no
long-range correlation between the $S_j$ variables either since
they are just local sums of the $\sigma_i$'s. However, a hidden
{\em string order} emerges in the system as the temperature
$T\rightarrow 0$. This can be seen from the string variable
\begin{equation}
L_{MN}=\sum_{i=M}^NS_i=\sum_{i=2M}^{2N+1}\sigma_i.
\label{L_N}
\end{equation}
 At $T=0$, the nearest neighbor spins $\sigma_{2j-1}$ and $\sigma_{2j}$ form $\pm 1/2$ pairs and sum to zero except those
 on the first site $2M$ and the last site $2N+1$. Therefore the only possible values $L_{MN}$ can take are $L_{MN}=-1,0,1$,
 independent of the site indices $N$ and $M$. To satisfy this constraint, the corresponding $S_j$ variables must display
 the following string topological order: starting with an allowed chain of variables $\{S_j\}$, we create another
 (shorter) chain by removing all the $S_j$'s with $S_j=0$. The remaining chain, which is an Ising chain with $S_j=\pm1$
 only, displays perfect antiferromagnetic Ising order\cite{aklt}. The string order is a direct consequence of the
 non-trivial topology of the phase space for the $S_j$ variables at $T=0$. For an open chain of $2N+2$ sites from $i=0$
 to $i=2N+1$, the allowed states, in terms of the $S_j$'s, fall into four disconnected topological sectors labelled by the
 boundary spin variables $\sigma_0,\sigma_{2N+1}$. States in a given sector can be converted to each other through the
 spin exchange within each pair of sites $(2j-1,2j)$, but conversion between states in different topological sectors must
 involve the flip of boundary spins. The string topological order is responsible for the appearance of fractionalized
 $s=\pm 1/2$ Ising spins at the two ends of the chain when $\{S_j\}$ are considered to be the fundamental variables.

\begin{figure}[htbp]
\includegraphics[width=2.7in]{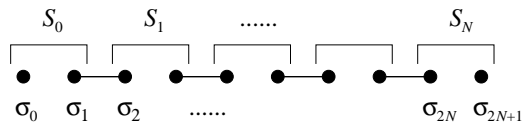}
\caption{The disjoint AKLT chain. Horizontal bonds indicate the nearest neighbor
coupling of Eq. (\ref{H_s_1D}).
}
\label{fig1}
\end{figure}

\section{The antiferromagnetic Ising model on a triangular lattice}
 We next consider a more realistic and non-trivial example of the topological glass state: the classical
 antiferromagnetic Ising model on a two-dimensional (2D) triangular lattice\cite{afi}. The model is defined by the
 Hamiltonian
\begin{equation}
\label{ising}
H=J\sum_{\langle i,j\rangle}S_iS_j+\sum_{i}h'(S_i;S),
\end{equation}
 where $S_i=\pm1$ and $\langle i,j\rangle$ denotes nearest neighbor lattice site pairs. The second term $h'(S_i;S)$ on
 the right-hand-side denotes all other plausible short range interactions such as the second and third nearest neighbor
 interactions, magnetic field, etc. The model has been well studied by many
 authors\cite{afi,blote1,dimer,rg,blote3,blote2,elser}. We shall first review briefly properties of the ground state at
 $h'=0$, and then examine how topology dictates properties of the system at low temperatures with $J\gg h',k_BT$.

\section{Ground state and topology of the reduced phase space}
 Due to the geometrical frustration associated with the triangular lattice, the ground state of (\ref{ising}) at $h'=0$ is
 not unique. Instead, for three spins on a triangle, the energy is minimized if the spins are not all parallel and there
 are $2\times3=6$ degenerate configurations. Extending the analysis to the whole lattice, we see that any spin
 configuration that satisfies the constraint
\begin{equation}
\sum_{i=1,2,3}S_i=\pm 1
\label{constraint}
\end{equation}
on every triangle in the lattice minimizes the energy. Equation (\ref{constraint}) defines a {\it reduced phase space}
 for the spin configurations. Since the number of such configurations grows exponentially with the system size, there is
 an extensive ground state degeneracy in this model\cite{afi}.

 The constraint (\ref{constraint}) encodes a topological order which can be illustrated in a number of equivalent ways,
 including the tiling representation, the string representation, the Solid-On-Solid (SOS) representation\cite{blote1} and
 the dimer representation on the dual lattice\cite{dimer}. The tiling, string and SOS representations are sketched in
 Fig. 2. The number of rhombuses in each of the three orientations, or the number and orientation of the unbroken strings,
 are topological properties of a given spin configuration which cannot be changed via single spin flips or spin flips in
 a localized region. These properties can be equivalently stated in the SOS representation described
 below which is more convenient for analytical treatment.

\begin{figure}[htbp]
\includegraphics[width=2.5in]{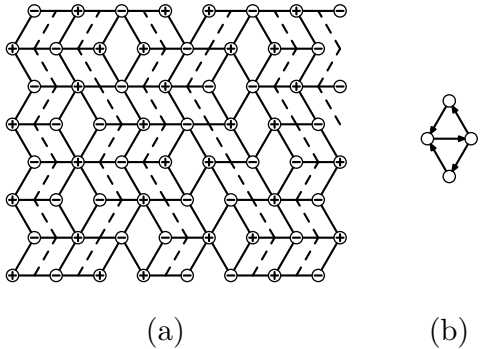}
\caption{Tiling, string and SOS representation.(a) Connecting neighboring spins that are antiparallel results in the
 tiling representation (solid lines). The string representation is given by connecting the midpoints of the horizontal
 edges of the rhombuses (dashed lines). (b) The directions $\hat{s}_{ij}$ of nearest neighbor bonds in the triangular
 lattice used to define height increments in the SOS representation.} \label{fig2}
\end{figure}

 In the SOS representation, a ``height" variable $h_i$ is assigned to every lattice site $i$ according to the following
 rule: first we assign a direction $\hat{s}_{ij}$ to all nearest neighbor bonds such that they point in counter-clockwise
 (clockwise) directions when travelling through up (down) triangles in the lattice [see Fig. 1(b)]. Starting from a site
 $i=0$ with spin $S_0$ and (arbitrarily assigned) height $h_0$, we travel to the next site $i=1$ along $\hat{s}_{01}$. A
 height $h_1=h_0-2$ is assign to site $1$ if the spin $S_1$ on the site is equal to $S_0$, otherwise $h_1=h_0+1$. The
 construction is then repeated to the third site. Since Eq. (\ref{constraint}) enforces two antiparallel and one parallel
 pairs of spins on every triangle, a consistent way of assigning the heights is obtained. The procedure can be continued
 to all sites on the lattice to yield a consistent and unique set of heights from a given spin configuration.
 Mathematically, the height $h_i$ at any given site $i$ is related to $h_0$ by
 \begin{subequations}
 \label{sh}
 \begin{equation}
 \label{height}
 h_i=h_0-{1\over2}\sum_{j\in L}(\hat{l}_{j,j-1}.\hat{s}_{j,j-1})\left(3S_jS_{j-1}+1\right)
 \end{equation}
 where $L$ denotes a chosen path on the lattice from site $0$ to $i$ and $j=1,2,...,i$ are sites on the path $L$.
 $\hat{l}_{j,j-1}$ is a unit vector pointing from site $j-1$ to site $j$. Notice that the choice of path $L$ is completely
 arbitrary. Conversely, a unique spin configuration can be constructed from a given height pattern starting from a given
 spin $S_0$ at site $0$. The spin at site $i$ is related to $S_0$ by
 \begin{equation}
 \label{spinp}
 S_i=e^{i\pi\sum_{j\in L}(h_j-h_{j-1})}S_0=e^{i\pi(h_i-h_0)}S_0.
 \end{equation}
 \end{subequations}
Equation (\ref{sh}) defines a one-to-one mapping between $\{S_i\}$ and $\{h_i\}$
apart from an overall height shift $h_0$.

%The last equality follows from the single-valuedness of the $h_i$'s and reflects the fact
%that the spin value $S_i$ is independent of the path $L$ chosen.

To show that the reduced phase space breaks into topologically distinct
sectors, let us consider the model on a torus (i.e., periodic boundary conditions).
For a system of size $L_x\times L_y$, where
$x$ is along one of the $\hat{s}_{ij}$ directions [Fig. 2(b)],
the boundary condition is $S_{x,L_y}=S_{x,0}$ and $S_{L_x,y}=S_{0,y}$.
In terms of the height variables, however, the boundary condition takes the form
$h_{x,L_y}=h_{x,0}+2n$ and $h_{L_x,y}=h_{0,y}-2L_x+3m$,
where $n$ and $m$ are integers determined by the
spin configuration and lie in a certain range set by $L_y$ and $L_x$.
Spin configurations with different values of $(n,m)$ belong to different
topological sectors because any set of local spin flips consistent with (\ref{constraint})
do not modify height increments across the system.

\section{Topological excitations and topological phase transitions}
 We next consider low temperatures $k_BT\ll J$. In this case, excitations with all three spins on a triangle pointing at
 the same direction are allowed with an excitation energy $2J$. This excitation is called a vortex and has a non-trivial
 topological structure. In the SOS representation, presence of a vortex makes it impossible to assign heights $h_i$'s in
 a consistent way. A discontinuity $\Delta h=6q$ is found for the $h_i$ variables on the triangle enclosing the vortex
 [$q=+(-)1$ for vortices located on up(down) triangles = charge of vortex]. A line of discontinuity (i.e., branch cut),
 extending from the center of the vortex to infinity or to the center of another vortex with opposite charge, is needed
 to restore a unique height configuration (see Fig. 3 for an example). Notice that $e^{i6\pi}=1$ and there is no
 uncertainty in the corresponding spin assignment. The topological nature of vortex excitation can also be seen in the
 string representation, which represents the point of annihilation of two strings\cite{blote2} resulting in a difference
 in incoming string number (from top of lattice) and outgoing string number (to bottom of lattice) by two (see Fig. 3).

 \begin{figure}[htbp]
 \includegraphics[width=2.0in]{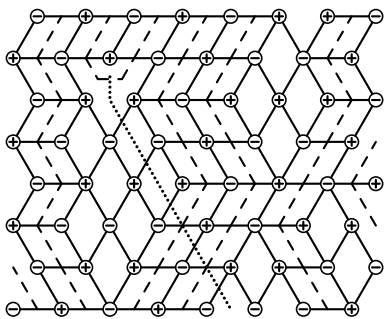}
 \caption{A vortex in tiling, string and SOS representation (an elementary triangle with identical spins at three
 corners). A singular line (dotted line) extends from the center of the vortex to infinity (or another vortex).
 The heights at the two sides of the singular line have a discontinuity $\Delta h=6$.}
 \label{fig3}
 \end{figure}

 Despite the extensive ground state entropy and the discrete nature of the spins, the topological order resulting from
 the constraint (\ref{constraint}) is quite analogous to that of the XY model defined by a phase field $\theta(\vec{x})$
 on the lattice. In fact, one may relate the local phase gradient $\vec{\nabla}\theta$ in the XY model to the local
 height gradient for neighboring sites,
 \begin{equation}
 \vec{A}_{j,j-1}=-{1\over 2}\hat{s}_{j,j-1}(3S_jS_{j-1}+1).
 \label{A-vec}
 \end{equation}
 In terms of $\vec{A}_{j,j-1}$, the flux through a closed contour $L$ on the lattice is given by,
 \begin{equation}
 \label{torder}
 F_L=\sum_{j\in L}\hat{l}_{j,j-1}\cdot\vec{A}_{j,j-1}=6q,
 \end{equation}
 where $q=0,\pm 1,\ldots$ is the net number of vortices enclosed by $L$. The reduced phase space at $T=0$ has $F_L=0$
 for any $L$. A more relaxed definition of topological order which applies also at nonzero temperatures is to demand
 $F_L\simeq 0$ for sufficiently large $L$, i.e., the vortices are in the confined phase.

 Using the XY model as an analogy, we can now understand the low-energy, large-distance properties of (\ref{ising}) and
 the nature of the transition to the disordered phase in terms of vortex excitations that form
 a dilute gas at low temperatures. The statistics of vortices is described by an effective Coulomb gas model\cite{blote3},
 \begin{equation}
 \label{cg}
 -{H_{\rm CG}\over k_BT}={1\over 2}\sum_{i\neq j}q_iq_jG(\vec{r}_i-\vec{r}_j)+\mu\sum_iq_i^2,
 \end{equation}
 where $G(r)\simeq g\ln (r/a)$ ($a$ = lattice spacing), $q_i$ is the vortex charge, and $\mu\simeq -2J/k_BT$. For $h'=0$,
 it is found that\cite{blote3,blote2} the ground state entropy yields $g=2$ which is less than the critical value $g_c=4$
 for vortex confinement in the zero fugacity limit. In this case, the vortex gas is in the disordered phase at any finite
 temperature $T>0$ ($\mu$ finite). A vortex confined phase can be stabilized by introducing a weak next-nearest neighbor
 ferromagnetic coupling $h'=-J_1\sum_{(i,j)\in {\rm n.n.n}}S_iS_j$ which reduces spin 
(or height) fluctuations and pushes $g$ to much higher
 values as $T\rightarrow 0$. Consequently, a Kosterlitz-Thouless transition to the topologically ordered phase
 takes place at $k_BT_{\rm KT}\simeq 8.48J_1\ll J$, followed by a second transition into a (ordered) flat phase at a
 lower temperature $k_BT_R\simeq 4.42J_1$\cite{j1a}. Spin-spin correlation changes from an exponential decay
 to a power-law decay at $T_{KT}$ without breaking of symmetry.

%Topological phase transition occurs when the Hamiltonian changes in such a way that the structure of the low
  %energy reduced phase space in the model changes. The following spin model provides an example of topological
  %phase transition. The model is a "double-layer" generalization of the nearest neighbor Ising model
  %on triangular lattice with spins $S_{1i}$ and $S_{2i}$ at the two layers respectively. The energy function is
  %\begin{equation}
  %\label{h2}
  %H=J\sum_{<i,j>}(S_{1i}S_{1j}+S_{2i}S_{2j})+J'\sum_iS_{1i}S_{2i}.
  %\end{equation}
  %When $J'>(<)0$, the ground state of the system is characterized by a single layer Ising model with
  %effective spin $S_{a(s)i}={1\over2}(S_{1i}-(+)S_{2i})$ with energy function
  %\[
  %E=2J\sum_{<i,j>}S_{\beta i}S_{\beta j}-N|J'|,  \]
  %where $N=$ total number of lattice sites and $\beta=s(a)$ when $J'<(>)0$. The topological order is
  %characterized by Eq.\ (\ref{torder}) with $F_L\rightarrow F_L^{\beta}$ and
  %\begin{equation}
  %\label{torder2}
  %\vec{A}_{j,j-1}\rightarrow\vec{A}^{\beta}_{j,j-1}=\hat{s}_{j,j-1}(3S_{\beta j}S_{\beta j-1}+1).
  %\end{equation}
  %The anti-symmetric to symmetric spin-binding transition from $J'>0$ to $J'<0$ is also a topological phase
  %transition. At both sides of the transition the spin system is disordered but the change can be detected by
  %different behaviors of the topological order parameters $F_L^s$ and $F_L^a$.

Additional terms in the Hamiltonian $h'$ do not affect the topological nature of the
low-temperature phase as long as they are smaller than $J$ and $J_1$ such that
vortices remain confined. In this case $h'$ can be expressed in the
SOS representation using Eq.\ (\ref{spinp}). Its main effect is to create different
phases in the system\cite{blote2,j1a,j1b} characterized by different behaviors of
the height variables $h_i$, meaning that $h_i$ has replaced the spin variables
to become an effective local order-parameter.

 \section{Higher dimensions}
The XY-type topological order exhibited by spin configurations in the reduced phase 
space defined by Eq. (\ref{constraint}) can be extended to higher dimensions.
The scheme of Eq. (\ref{A-vec}) assigns local height gradients
according to the nearest-neighbor bonding situation, together with a sign convention
set by the bond direction vectors $\hat{s}_{i,j}$.
A consistent height assignment for all sites can be made if and only if the zero-flux
condition $F_L=0$ is satisfied for every closed contour on the lattice. 
Quite generally, $F_L$ can be decomposed into contributions from 
``elementary plaquettes'' of the lattice, which are triangles in the 2D example. 
Therefore, if $F_L=0$ holds for every elementary plaquette, then a topological order 
of the XY-type is established.

 As an example, let us consider the three-state chiral Potts model on a body-centered-cubic (bcc) lattice defined by the
 Hamiltonian\cite{forrest}
 \begin{equation}
 H_{\rm P} =-\sum_{i,\alpha}\delta\{\sigma_i+1,\sigma_{i+\hat{s}_\alpha}\},
 \label{cPotts}
 \end{equation}
 where $\sigma_i=0, 1, 2$, $\hat{s}_\alpha$ is one of four nearest neighbor vectors pointing from the center to the four
 non-adjacent corners of the cube, and $\delta\{a,b\}=1$ if $a=b$ (mod 3) and 0 otherwise. 
The elementary plaquettes of the bcc lattice are formed by the four vectors
$\hat{s}_\alpha$ linked head-to-tail in six alternative sequences.
Since each such plaquette contains four sites, at least one of the four bonds is frustrated
under the energy (\ref{cPotts}). It has been shown in Ref.\cite{forrest} that the
ground state of (\ref{cPotts}) contains precisely one frustrated bond per elementary
plaquette and an extensive entropy as in the 2D case. 
Assigning a height difference of 1 for the satisfied bonds and -3 for the frustrated
bonds along directions defined by the $\hat{s}_\alpha$'s, we obtain a unique height
configuration $\{h_i\}$ from a given Potts spin configuration $\{\sigma_i\}$, apart
from an overall height shift. The inverse mapping is given by $\sigma_i=h_i$ (mod 3).
With the help of the height variables, various topological sectors of the ground state 
can be identified in a similar way as in the 2D case.

\section{Summary}
Summarizing, we propose in this paper a new kind of glassy state in
classical systems called topological glass which are analogies of topological order
in quantum systems. The existence of these states are illustrated by two explicit
examples. Topological order can emerge in classical systems at
low temperature as a result of the non-trivial topology of the
reduced phase space. In our example of the AKLT state, topological order in the
classical model is directly linked to the corresponding quantum state.
The topological order in the 2D antiferromagnetic Ising model on the triangular
lattice is of the XY-type which has high dimensional counterparts.
It would be interesting to explore topological order associated with other symmetries 
such as those discussed in Ref.\cite{sethna}. 
Our work is only the first step towards bridging the study of
topological order in quantum and classical systems. 

\acknowledgements
The authors acknowledge support from the Research Grants Council of the HKSAR
through grant CA05/06.SC04. TKN acknowledges the hospitality of the
Kavli Institute for Theoretical Physics China at which this work started.


\begin{thebibliography}{0}
\bibitem{fqhe} S.C. Zhang, Int. J. Mod. Phys. {\bf B6}, 25 (1992).
\bibitem{wen} see for example, X.G. Wen, {\em Quantum Field Theory of Many-Body Systems},
Oxford University Press 2004).
\bibitem{fulde} F. Pollmann, J.J. Betouras, K. Shtengel and P. Fulde, Phys. Rev. Lett. {\bf 97},
170407 (2006).
\bibitem{kitaev} A. Kitaev, Ann. Phys. {\bf 321}, 2 (2006).
\bibitem{tao} X.-Y. Feng, G.-M. Zhang and T. Xiang, Phys. Rev. Lett. {\bf 98}, 087204 (2007).
\bibitem{aklt} I. Affleck, T. Kennedy, E.H. Lieb abd H. Tasaki, Phys. Rev. Lett. {\bf 59},
799 (1987).
\bibitem{afi} R.M.F. Houtappel, {\em Physica} {\bf 16}, 425 (1950).
\bibitem{blote1} H.W.J. Bl$\ddot{o}$te and H.J. Hilhorst, J. Phys. A {\bf 15}, L631 (1982).
\bibitem{dimer} F.Y. Wu, Phys. Rev. {\bf 168} 539 (1968).
\bibitem{rg} J.V. Jos$\acute{e}$, L.P. Kadanoff, S. Kirkpatrick and D. Nelson,
Phys. Rev. B {\bf 16}, 1217 (1977); H.J.F. Knops, Phys. Rev. Lett. {\bf 39}, 766 (1977).
\bibitem{blote3} H.W.J. Bl$\ddot{o}$te and M.P. Nightingale, Phys. Rev. B {\bf 47}, 15046 (1993).
\bibitem{blote2} B. Nienhuis, H.J. Hilhorst and H.W.J. Bl$\ddot{o}$te, J. Phys. A{\bf 17},
3559 (1984).
\bibitem{elser} V. Elser, J. Phys. A{\bf 17}, 1509 (1984).
\bibitem{j1a} X.-F. Qian and H.W.J. Bl$\ddot{o}$te, Phys. Rev. E {\bf 70}, 036112 (2004).
\bibitem{j1b} S. Fujiki, K. Shutoh, Y. Abe and S. Katsura, J. Phys. Soc. Japan. {\bf 52},
1531 (1983); H. Takayama, K. Matsumoto, H. Kawahara, and K. Wada, {\it ibid.}
{\bf 52}, 2888 (1983); D.P. Landau, Phys. Rev. B {\bf 27}, 5604 (1983).
\bibitem{forrest} B.M. Forrest and L.-H. Tang, Phys. Rev. Lett. {\bf 64}, 1405 (1990).
\bibitem{sethna} J. Sethna, in {\it 1991 Lectures in Complex Systems}, Eds. L. Nagel and D. Stein,
Santa Fe Institute Studies in the Sciences of Complexity, Proc. Vol. XV, Addison-Wesley, 1992.
\end{thebibliography}
\end{document}